\documentstyle[preprint,prc,aps]{revtex}
\begin{document}
\draft

\title{Calibration of Plastic Phoswich Detectors for Charged Particle
Detection}
\author{D. Fox, D. R. Bowman, G. C. Ball,
A. Galindo-Uribarri, E. Hagberg, and D. Horn}
\address{AECL, Chalk River Laboratories, Chalk River, Ontario,
K0J 1J0, Canada}
\author{L. Beaulieu and Y. Larochelle}
\address{Laboratoire de Physique Nucl\'eaire, Universit\'e Laval, Ste-Foy,
Qu\'ebec, G1K 7P4, Canada}
\date{\today}

\maketitle

\begin{abstract}
The response of an array of plastic phoswich detectors to ions of
$1\le Z\le 18$ has been measured from $E/A$=12 to 72 MeV.
The detector response has
been parameterized by a three parameter fit which includes both
quenching and high energy delta-ray effects.  The fits have a
mean variation of $\le 4\%$ with respect to the data.

\end{abstract}

\newpage

\section{Introduction}
\label{sec-intro}
In recent years a number of large arrays using plastic and/or CsI
scintillators have been built for heavy ion experiments
\cite{plasticball,4pi,flyseye,pouliot88,pagoda,dwarf,miniball,pruneau90,%
leegte92,cebra92,medea,alf,iori93,isis,indra1,indra2}.
While
these arrays offer a relatively inexpensive way of covering large solid
angles,
the nonlinear response of various scintillators as a function of the incident
particle's energy, charge, and mass poses a
challenge in their calibration.  To understand the nonlinear response of
various scintillators, calibration data covering a wide range of
projectile energy, charge and mass are required.
These requirements, combined with
the large number of elements in an array, make it virtually impossible to
do a detailed calibration after each experiment.

Thus a procedure by which the calibration of large scintillator arrays
can be achieved during or after an experiment with only a few
calibration points would be valuable.
The essential ingredient in such a procedure is
an understanding of the scintillator response to a wide variety and energy of
incident particles.
Considerable work has been done in the past to understand the complex
scintillator
response.  Birks proposed \cite{birks} a simple expression for the light
output which took into account the quenching of light output for cases of
very large energy loss.  Other authors proposed expressions which also took
into account high energy electrons, or ``delta rays'',
which escape from the primary ionization
column \cite{meyer,voltz,tarle}, and recently a new model of
light production by
secondary electrons has been suggested \cite{mich1,mich2}.
While some authors have applied these models to their data
\cite{badhwar,mcmahan}, many others have used empirical expressions
which may not be generally applicable
\cite{pouliot88,dwarf,leegte92,cebra92,newman,buenerd,becchetti,%
twenhofel,colonna,yves}.

In the present work we extend our previously published results
for CsI response \cite{horn} by incorporating
delta rays into the Birks' formalism and
applying the resulting expression to calibration data taken with plastic
phoswich detectors from the Chalk River/Laval forward array \cite{pruneau90}.
Details of
the calibration data are given in the next section.
In Section~\ref{sec-param}
the parameterization of the light output is derived.  Results with
fits to the parameterization of Section~\ref{sec-param} to the calibration
data are discussed in
Section~\ref{sec-results}.
A step by step description of the procedure used to apply the calibration
to data is described in Section~\ref{sec-apply}.
Conclusions from the present work are given
in Section~\ref{sec-conclusions}.

\section{Measurements}
\label{sec-meas}

The CRL/Laval forward array phoswich detectors \cite{pruneau90}
consist of 0.7 mm thick fast plastic $\Delta E$ (BC408 from Bicron
Corporation) elements
heat-pressed to 76 mm thick slow plastic $E$ (BC444) elements.
The photomultiplier
signal is split, and integrated with two charge-to-digital converters (QDCs).
The first QDC, with a gate width of 57 ns, integrates the early portion of
the signal which corresponds primarily to energy lost in the fast
plastic $\Delta E$ counter.  The second QDC, with a gate width of 183 ns is
delayed by 210 ns with respect to the first gate.  This QDC integrates
a later portion
of the signal which corresponds primarily to the energy deposited in the
slow plastic $E$ counter.

An example of a $\Delta E-E$ particle identification plot is shown in
Fig. \ref{fig-de-e}.  Ions which stop in the fast plastic $\Delta E$ lie
along the steeply rising line, or ``backbone'', near the y-axis.
The deviation of the backbone
from the vertical is due to the tail of the fast plastic signal which
extends into
the delayed slow gate.  Neutral particles, almost all of which interact only
in the slow plastic $E$, are visible in the inset in Fig. \ref{fig-de-e}
as the straight line below the hydrogen isotopes.  The nonzero slope of
the neutral line is
due to the leading edge of the slow signal which lies within the fast gate.

In order to measure the response of phoswich detectors from the
CRL/Laval forward array, four elements of the array
were mounted on a movable trolley which allowed them to be placed directly
in the beam.
The detectors were exposed to a series of secondary beams
produced from an $^{18}$O beam at $E/A$=36.5 MeV incident on two
$^{12}$C production targets at the exit of the cyclotron at the TASCC
facility at Chalk River.
Six beamline rigidity
settings, listed in Table~\ref{tab-beams}, were used to select different
energies and isotopes.
The production target-beamline rigidity combinations
produced up to 94 calibration points per detector.  A total of 41 different
isotopes with $1\le Z\le10$ and $12.0\le E/A\le72.6$ MeV were produced.
Representative $\Delta E-E$ plots
are shown in Fig.~\ref{fig-data} for two beam rigidities.
Only calibration points
with more than ten counts were used in the subsequent fitting procedure.

In order to convert the centroids of the calibration points
into values proportional to the light
emitted in each element of the phoswich, the QDC pedestals
must be subtracted, and
a correction must be applied for the cross talk of the $\Delta E$ and $E$
signals in the two QDC gates.
The correction for the latter signal
can be determined from the slopes and intercepts
of the backbone and neutral lines.  We take

\begin{equation}
s=L+m_bdL+b_b
\label{eq-lmeas}
\end{equation}
\noindent{ and}
\begin{equation}
f=dL+m_nL+b_n,
\label{eq-dlmeas}
\end{equation}
\noindent{where $s$ and $f$ are the measured $E$ (slow) and $\Delta E$ (fast)
signals, $L$ and $dL$ are the ``true''
$E$ and $\Delta E$ signals,
$m_n$ ($m_b$) is the slope (inverse slope) of the neutral line (backbone),
and $b_b$ and $b_n$ are
the QDC pedestals.  Solving for $dL$ the measured signals can now be
related to $L$ by:}

\begin{equation}
L={ s-m_bf+m_bb_n-b_b \over 1-m_bm_n}.
\label{eq-ltrue}
\end{equation}

In this paper we describe only the calibration of the slow plastic signal.
The total energy may then be obtained by determining the energy loss in
the $\Delta E$ element from a table lookup based on the
ion's element number and the energy deposited in the slow plastic.

\section{Light Output Parameterization}
\label{sec-param}
In order to account for the observed nonlinearity of scintillator
light output Birks proposed a simple expression for the
differential light output ${dL \over dx}$
in terms of a quenching coefficient $kB$ times the energy loss
${dE \over dx}$:

\begin{equation}
{ dL \over dx} ={S \over 1+kB {dE \over dx}}{dE \over dx},
\label{eq-birks}
\end{equation}

{\noindent
where $S$ is the scintillation constant \cite{birks}.  Under this formulation,
an increasing energy loss leads to greater quenching as
the primary ionization column becomes saturated.  The next step is to consider
the effect of high energy electrons, or ``delta rays'',
which have sufficient energy to escape from the
primary ionization column.
These delta rays, whose scintillation efficiency is nearly 100\% \cite{meyer},
will begin to dominate the light output for very heavy ions where the
light output within the primary ionization column is almost totally quenched.
If $F_s$ is the fraction of the energy carried by delta rays then
the differential light output may be expressed as}

\begin{equation}
{ dL \over dx} ={S(1-F_s) \over 1+(1-F_s)kB {dE \over dx}}{dE \over dx}+
SF_s{dE \over dx}.
\label{eq-birks2}
\end{equation}

\noindent{The fraction $F_s$ may be expressed as \cite{voltz,ahlen}}

\begin{equation}
F_s={1 \over 2}{\ln(2m_ec^2\gamma^2\beta^2/T_0)-\beta^2 \over
\ln(2m_ec^2\gamma^2\beta^2/I)-\beta^2}.
\label{eq-fs}
\end{equation}

\noindent{where $\beta$ and $\gamma$ are calculated from
the ion velocity, $m_e$
is the electron rest mass, $T_0$ is the minimum
electron energy needed to escape
from the primary ionization column, and $I$ is the ionization potential of the
scintillator, $I\approx$0.048 keV for plastic scintillator.
At intermediate energies $\gamma^2\approx$1, $\beta^2\ll 1$, and
Eq. (\ref{eq-fs}) simplifies to:}

\begin{equation}
F_s={1\over 2}\left( 1- {\ln(T_0/I) \over \ln\left({a\over I}E/A\right)}\right)
\label{eq-fs2}
\end{equation}
\noindent{where $a={4m_e \over m_0}$ and $m_0$ is the nucleon rest mass.
Since $F_s$ is negative for $E/A<{T_0 \over a}$, Eq.
(\ref{eq-birks2}) is integrated up to $E/A={T_0\over a}$ with $F_s=0$.
The light output may now be expressed as:}

\begin{equation}
L=SE-SkB\int_0^{T_0 \over a} {{dE \over dx} \over 1+kB{dE\over dx}} dE-
S{kB \over 2}\int_{T_0\over a}^E {(1+R)^2{dE \over dx}
\over 2+kB(1+R){dE \over dx}} dE
\label{eq-lo}
\end{equation}

\noindent{where
$R={\ln(T_0/I) \over \ln\left({a\over I}E/A\right)}$.  There are three
parameters in this expression:
the scintillation constant $S$, the quenching factor
$kB$, and the electron kinetic energy cutoff $T_0$.
The quantity $S$ is measured in combination with an overall
gain factor from the readout device, but the latter two parameters are,
in principle, properties
of the scintillator material and should be the same for each detector.}

\section{Results}
\label{sec-results}
\subsection{$^{18}$O Calibration}
\label{sec-results-o18}
The free parameters in Eq.~(\ref{eq-lo}) were fitted to
the data described in Section~\ref{sec-meas}
for the energy deposited in the slow plastic.
The energy lost  in the fast plastic was calculated using the energy-loss
code STOPX \cite{stopx} and subtracted from the incident energy to give
the energy deposited in the slow plastic.
In order to obtain satisfactory fits it was necessary
to separately fit the light,$1\le Z\le 3$, and heavy, $4\le Z\le 10$, ions.
The values for the electron kinetic energy cutoff parameter $T_0$, from the
four detectors,
were found to be in good agreement for the light ion fit,
see Table~\ref{tab-li-res}.  The fit was
then repeated with $T_0$ fixed at the average value of 2.85 keV.  The
results of this second fit showed a better agreement in the four values
of $kB$ without significantly affecting the quality of the fit.  A final pass
was made with  $kB$ at the average value of 8.25.  Again, there was no
significant effect on the overall quality of the fit.
The mean difference between the fit and data is $\le 4\%$.
The light ion data and final fit results
are shown in Fig.~\ref{fig-li-res} for one detector.

In fitting the heavy ion data, $4\le Z\le10$,
to Eq.~(\ref{eq-lo}) it was necessary to
add a quadratic term to Eq. (\ref{eq-ltrue}) to account for a slight curvature
of the backbone for high $\Delta E$.  Equation~\ref{eq-ltrue} becomes

\begin{equation}
L={ s-m_bf+m_bb_n-b_b \over 1-m_bm_n}-qf^2.
\label{eq-ltrue2}
\end{equation}

\noindent{The additional parameter $q$ may be included in the fit,
or determined independently by making the
backbone pass through a punch-in point of a high $Z$ line such as the one
indicated in Fig.~\ref{fig-de-e}.
Once again, the values for $T_0$ and $kB$ were found to be
in good agreement from detector to detector, see Table~\ref{tab-hi-res}.
The mean difference between the fit and data is $\le 2\%$.
The results of the fits for
the heavy-ion data are shown Figure~\ref{fig-hi-res} for one detector.}

After fitting both the light- and heavy-ion data it was found that the ratio
of the light ion gain factor $S_{LI}$ to the heavy ion gain factor $S_{HI}$
was $\approx$1.30 for all four detectors; see Table~\ref{tab-ratio}.
The ability to obtain
a single detector-independent set of parameters $kB$ and $T_0$
combined with a fixed ratio for the overall gain parameters, permits
the calibration of other data sets with a single calibration point.

\subsection{Gate Tests}
\label{sec-results-gate}
The sensitivity of the gains to changes in the slow gate width
and delay were checked by separately varying the gate width
and delay from their
normal values.  Figure \ref{fig-gate-test}(a) shows the percentage shift
when the gate width was changed from the
usual value of 183 ns.
Reducing the gate width to 150 ns resulted in an
average shift of
$\approx-13.5\%$, with no apparent $Z$ dependence and less than a 1\% scatter
in the individual shifts.  Lengthening the gate width to 210 ns produced an
average shift of $\approx+11.5\%$ with a small $Z$ dependence.
The effect of varying the slow gate delay from the normal value of 210 ns
is shown in Fig.~\ref{fig-gate-test}(b).  For changes of $\pm$30 ns,
very small ($<\pm$1\%) $Z$ dependence was observed.  Reducing the
slow gate delay to 150 ns produced the largest observed $Z$ dependence.
Even in this case the observed $Z$ dependence is still small, $<\pm 2.5$\%,
with the largest shifts observed for Li and Be isotopes.

Data taken with phoswich detectors can be affected by differences in timing
between detectors.  Since timing differences result in a small
smear of slow gate
delays, the data in Fig.~\ref{fig-gate-test}(b) may be used to estimate
the effect on the measured energy of poor timing between phoswich detectors.
In the past we have experienced
timing differences of up to 10 ns for a few of the
phoswich detectors which has now been corrected.
{}From Fig.~\ref{fig-gate-test}(b) the effect of a 10 ns
time shift can be estimated to be less than 5\%.

\subsection{$^{37}$Cl Calibration}
\label{sec-results-cl37}
The extent to which the values of $kB$, $T_0$, and
$S_{LI}/S_{HI}$, which were obtained in Section~\ref{sec-results-o18},
are applicable to other data sets was checked by calibrating a
second data set with the parameter values obtained from the first data set.
In the second
data set, a single phoswich from the first ring of the array was placed
directly into a secondary beam ranging from Hydrogen to Argon ions
produced by a
$^{37}$Cl beam incident on a 50 mg/cm$^2$ C production target with a beamline
$B\rho$ setting of 1.584 Tm.  With $kB$
and $T_0$ fixed at 7.18 and 1.13 keV respectively, the data points for
$4\le Z\le 18$ were fit with just two free parameters, $S_{HI}$ and $q$.  The
fit results differed from the measured values by an average of 1.2\%.
The light ion data were then fit with $kB$=8.25 and $T_0$=2.85 keV and with
the light-ion gain factor, $S_{LI}$, fixed at $1.30S_{HI}$.
The average difference between the
predicted and measured centroids for $1\le Z\le 3$ was 5.6\%.  While somewhat
better fits were obtained when all the fit parameters were allowed to vary, the
calibrations obtained with the $^{18}$O fit parameters are quite satisfactory
for plastic phoswich detectors.

\section{Application to Data}
\label{sec-apply}
The parameterization described above in Eq.~\ref{eq-lo} is for the energy
deposited in the slow plastic $E$ detector.  In order to obtain the total
incident energy from this parameterization it is necessary to calculate the
energy deposited in the fast plastic $\Delta E$ detector.  Furthermore,
Eq.~\ref{eq-lo} can only be numerically inverted into a relation for energy as
a function of light.  The following steps are followed to
produce a table of the total incident energy as a function of the light
output of the slow plastic for a given ion:

1.  The energy loss $\delta E$ in the fast plastic $\Delta E$
detector is calculated for the incident fragment energy $E_{inc}$.

2.  Equation~\ref{eq-lo} is used to calculate the light output for the
energy deposited in the slow plastic $E_{slow}=E_{inc}-\delta E$.

3.  Steps 1 and 2 are repeated for a range of fragment energies.  The
results are used to build a table of light output as a function of
incident energy.

4.  The table in step 3 is inverted to give a table of the incident energy
as a function of the light output of the slow plastic.

5.  Steps 1 to 4 are repeated for each ion to which the calibration will
be applied.

Since isotopic resolution in the phoswichs is only acheived in the calibration
data through $B\rho$ separation, normal data sets use
calibrations assuming $A=2Z$ for $Z\ge2$ and
$A=1$ for $Z=1$.  This assumption introduces an error in the fragment
energy when the fragment has a different mass number than the assumed value.
For example the calculated energy will be $\approx$10\% less than the true
energy for $^8$Li and $\approx$5\% less for $^{18}$O.  For comparison the
slow plastic energy resolution is $\approx$5\%.

\section{Conclusions}
\label{sec-conclusions}

The response of plastic phoswich detectors from the CRL/Laval forward
array has been measured for a wide range of incident particles and energies.
The data are well reproduced by a fit based on the light output
relation of Birks with an additional term to include the effects
of high energy delta rays which escape from the primary ionization column.
The light output is characterized with just three parameters:
the gain factor $S$, the quenching factor $kB$, and an electron cut-off energy
$T_0$.  The quenching factor and electron cut-off energy have been found to
be constant from detector to detector.  In order to achieve reasonable fits
over a wide range of ions, $1\le Z\le 18$,
it was necessary to fit the light ($Z\le 3$) and heavy ($Z\ge4$)
ions separately.  The inability to fit all ions with the same set of
parameters may be due to pulse shape differences between the ions combined with
the integration of only part of the photomultiplier signal.
The mean difference between the fits and the data is
$\approx$4\% for light ions and $\approx$2\% for heavy ions.
The ratio of the gain factors for the separate
light- and heavy-ion fits has been found to be a constant.
It is thus possible to obtain future calibrations with a single
normalization point.
Tests demonstrated that changes in the QDC gates have little effect on
the calibrations other than in changing the overall gain factor.
The calibration procedure outlined in this paper has also been successfully
applied to CsI detectors over a wide range of incident particles from
$Z$=2 to 32.

\section{Acknowledgement}
This work was supported in part by the Natural Sciences and Engineering
Research Council of Canada.


\begin{figure}
\caption{$\Delta E$ (fast signal)$-E$ (slow signal) plot for
$^{70}$Ge+$^{nat}$Ti at
$E/A=35$ MeV at $\theta_{lab}=13.4^\circ$.  The inset shows an enlargement
of the lower left corner.}
\label{fig-de-e}
\end{figure}

\begin{figure}
\caption{$\Delta E$ (fast signal)$-E$ (slow signal)
plots for secondary beams for (a) $B\rho=1.50 Tm$
and (b) $B\rho=2.17 Tm$.  Selected isotopes are indicated.}
\label{fig-data}
\end{figure}

\begin{figure}
\caption{Final results of fits of Eq. (8) to the light ion data for
detector \#34.  The data are shown as points, the results of the
fits are indicated by lines.}
\label{fig-li-res}
\end{figure}

\begin{figure}
\caption{Selected fits of Eq. (8) to the heavy ion data for detector \#34.
The data are shown as points, the fit results are shown
as lines.}
\label{fig-hi-res}
\end{figure}

\begin{figure}
\caption{Sensitivity of the detector gains to changes in the slow gate
width and delay.  The data are from one detector for runs with $B\rho$=1.5 Tm.
Multiple isotopes are included for some fragment charges.
(a) Percentage shift in the detector gains for slow
gate widths of 150 and 215 ns relative to the normal slow gate width of 183 ns.
(b) Percentage shift in the detector gains for slow
gate delays of 150, 178, and 240 ns relative to the normal slow gate
delay of 210 ns.}
\label{fig-gate-test}
\end{figure}

\begin{table}
\caption{Beamline rigidity, $B\rho$, $^{12}$C production target thickness,
charge range of measured fragments, and degraded $^{18}$O energy.
For $B\rho$=2.17 Tm
the energy for fragments with $Z/A$=0.5 is given.}

\begin{tabular}{ddcd}
$B\rho$ & Production Target& $Z$ range & $^{18}$O Energy \\
(Tm)& (mg/cm$^2$) &      & (MeV/nucleon) \\ \tableline
1.25&         307 & 1-9  & 14.8 \\
1.33&         307 & 8    & 16.6 \\
1.50&          80 & 1-10 & 21.2 \\
1.75&          80 & 2-8  & 28.8 \\
1.86&          80 & 8    & 32.3 \\
2.17&          80 & 1-5  & 25.0 ($Z/A$=0.5) \\
\end{tabular}
\label{tab-beams}
\end{table}

\begin{table}
\caption{Light ion ($1\le Z \le 3$) fit parameters $T_0$ and $kB$,
and average percentage difference
between the fit and the slow-plastic data for each detector.
The average percentage difference includes only data points with 10 or more
counts which are at least 40 channels beyond the backbone.
The electron cut-off
energy $T_0$ was fixed after the first pass, and the quenching
coefficient $kB$ was fixed after the second pass.}
\begin{tabular}{c|ccc|ccc|ccc}
&\multicolumn{3}{c} {Pass 1}  & \multicolumn{3}{c} {Pass 2}
& \multicolumn{3}{c} {Pass 3} \\
Detector&$T_0$ & $kB$ &Difference& $T_0$ & $kB$ &Difference& $T_0$ & $kB$  &
Difference\\
        &(keV) &      &(\%)      &(keV)  &      &(\%)      &(keV)  &       &
 (\%) \\
        &      &      &          &(fixed)&      &          &(fixed)&(fixed)&
      \\
\tableline
33      &  2.84&  8.13& 3.9      &   2.85&  8.11& 3.9      &   2.85&   8.25&
3.9\\
34      &  3.24&  7.36& 3.0      &   2.85&  7.89& 2.8      &   2.85&   8.25&
2.9\\
35      &  2.71&  8.27& 3.2      &   2.85&  8.04& 3.3      &   2.85&   8.25&
3.3\\
48      &  2.60&  9.46& 2.2      &   2.85&  8.94& 2.4      &   2.85&   8.25&
2.4\\
\tableline
Average &  2.85&  8.31&          &       &  8.25&          &       &       & \\
\end{tabular}
\label{tab-li-res}
\end{table}

\begin{table}
\caption{Heavy ion ($4\le Z \le 10$) fit parameters, $T_0$ and $kB$,
and average percentage difference
between the fit and the slow-plastic data for each detector.
The average percentage difference includes only data points with 10 or more
counts which are at least 40 channels beyond the backbone.
The electron cut-off
energy $T_0$ was fixed after the first pass, and the quenching
coefficient $kB$ was fixed after the second pass.}
\begin{tabular}{c|ccc|ccc|ccc}
&\multicolumn{3}{c} {Pass 1}  & \multicolumn{3}{c} {Pass 2}
& \multicolumn{3}{c} {Pass 3} \\
Detector&$T_0$ & $kB$ &Difference& $T_0$ & $kB$ &Difference& $T_0$ & $kB$  &
Difference\\
        &(keV) &      &(\%) &(keV)  &      &(\%) &(keV)  &       & (\%) \\
        &      &      &     &(fixed)&      &     &(fixed)&(fixed)&      \\
\tableline
33      &  1.02&  7.18&  1.7&   1.13&  7.54&  1.8&   1.13&   7.18&1.8\\
34      &  1.31&  8.14&  1.5&   1.13&  7.51&  1.6&   1.13&   7.18&1.6\\
35      &  1.06&  6.98&  1.6&   1.13&  7.22&  1.7&   1.13&   7.18&1.7\\
48      &  1.13&  6.44&  1.6&   1.13&  6.43&  1.6&   1.13&   7.18&1.7\\
\tableline
Average &  1.13&  7.18&     &       &  7.18&     &       &       &\\
\end{tabular}
\label{tab-hi-res}
\end{table}

\begin{table}
\caption{Gain factors from light-ion, $S_{LI}$, and heavy-ion, $S_{HI}$,
fits to Eq. (8), and the ratio $S_{LI}/S_{HI}$.}
\begin{tabular}{cccc}
Detector & $S_{LI}$ & $S_{HI}$ & $S_{LI}/S_{HI}$\\ \tableline
       33&     4.54 &     3.52 & 1.29 \\
       34&     4.26 &     3.33 & 1.28 \\
       35&     5.11 &     3.94 & 1.30 \\
       48&     4.69 &     3.58 & 1.31 \\
Average  &          &          & 1.30 \\
\end{tabular}
\label{tab-ratio}
\end{table}

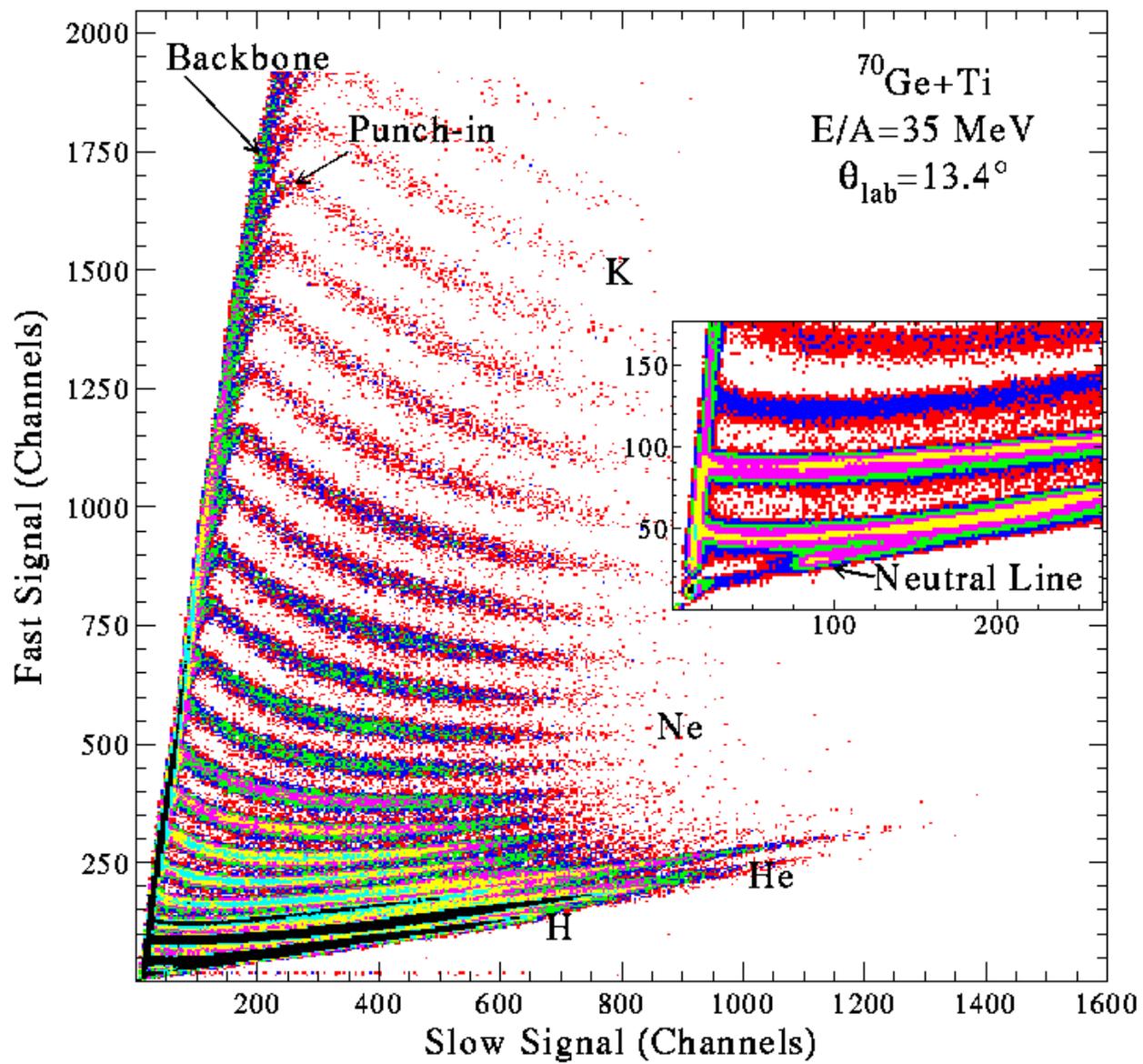
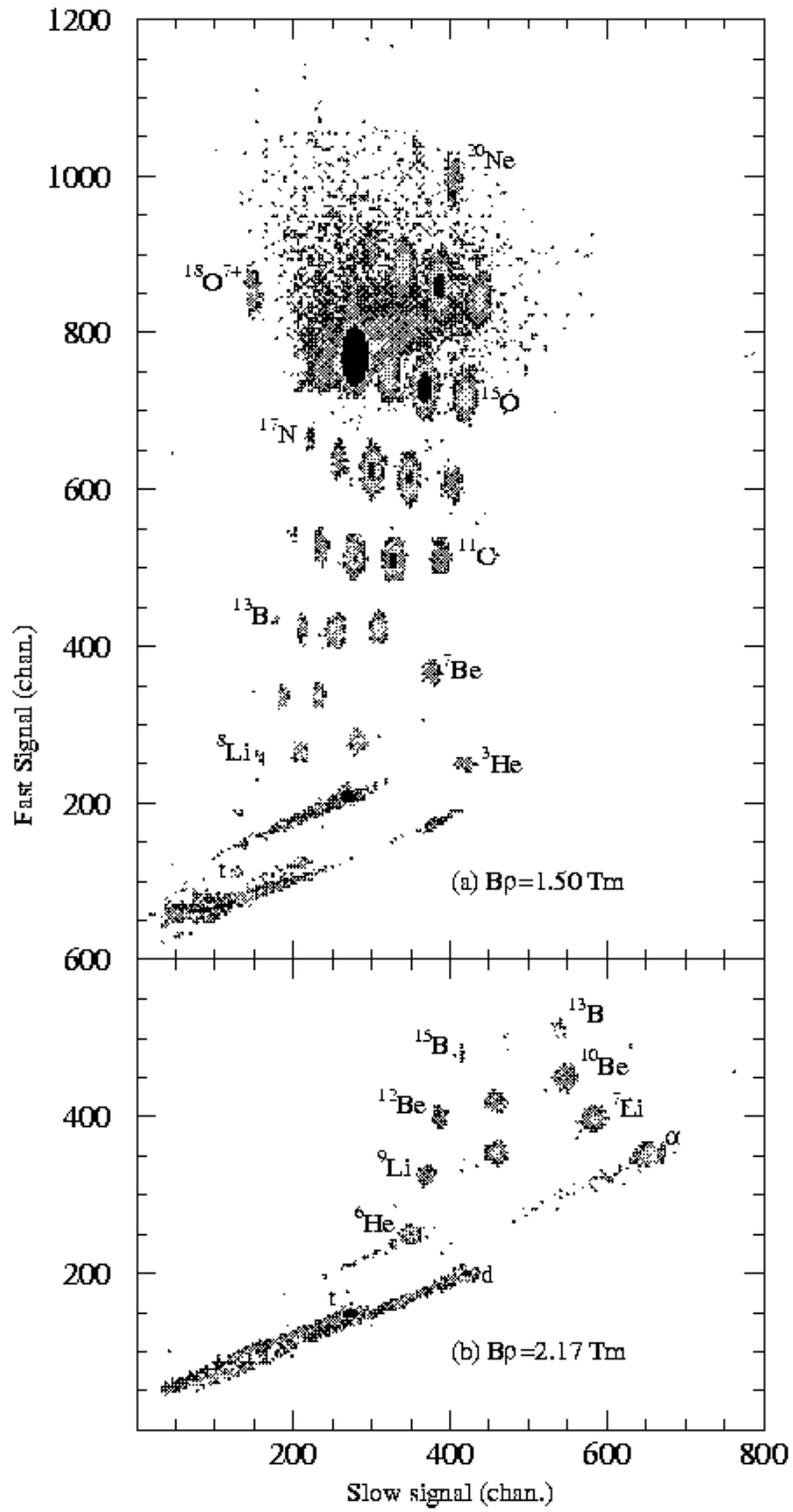
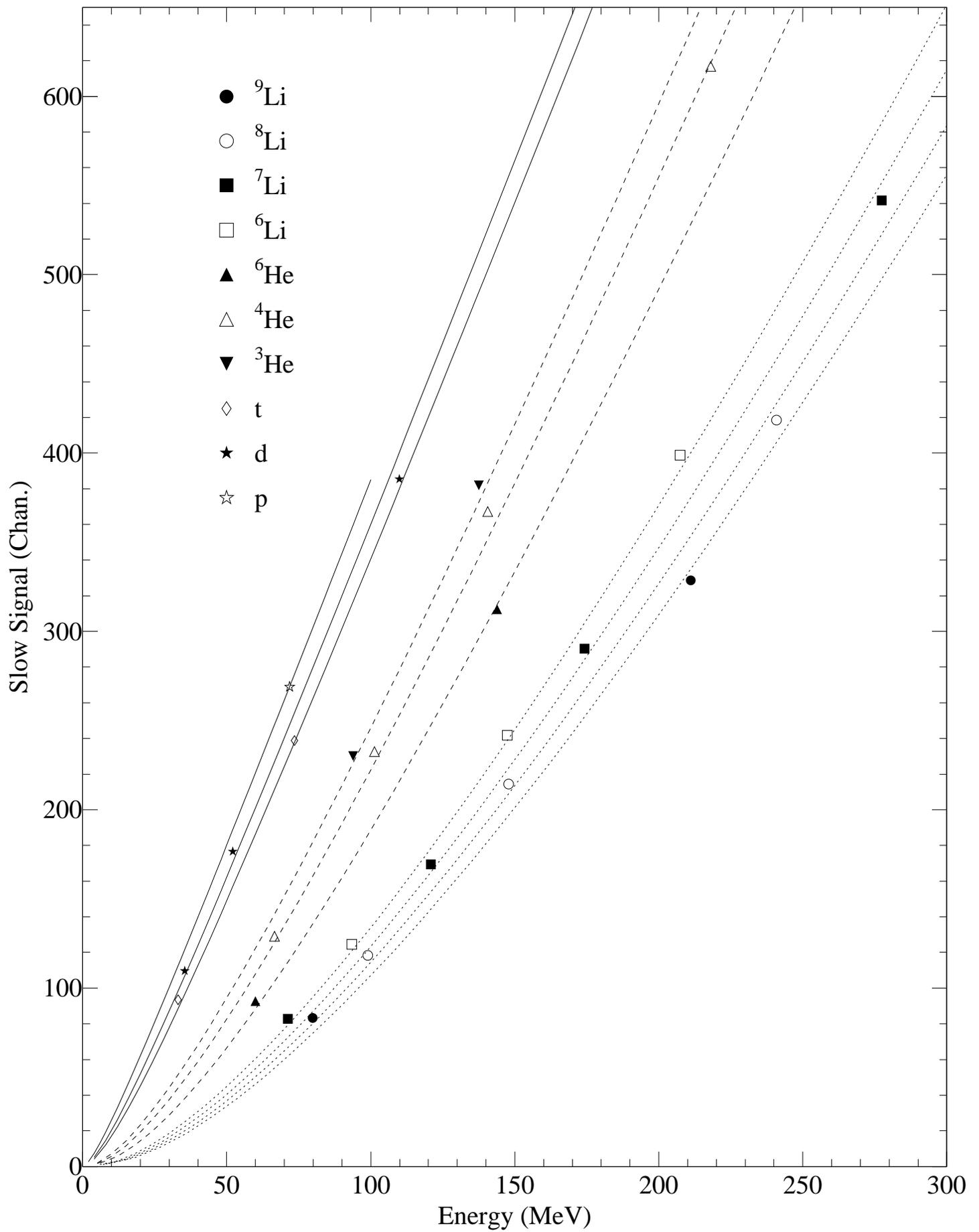
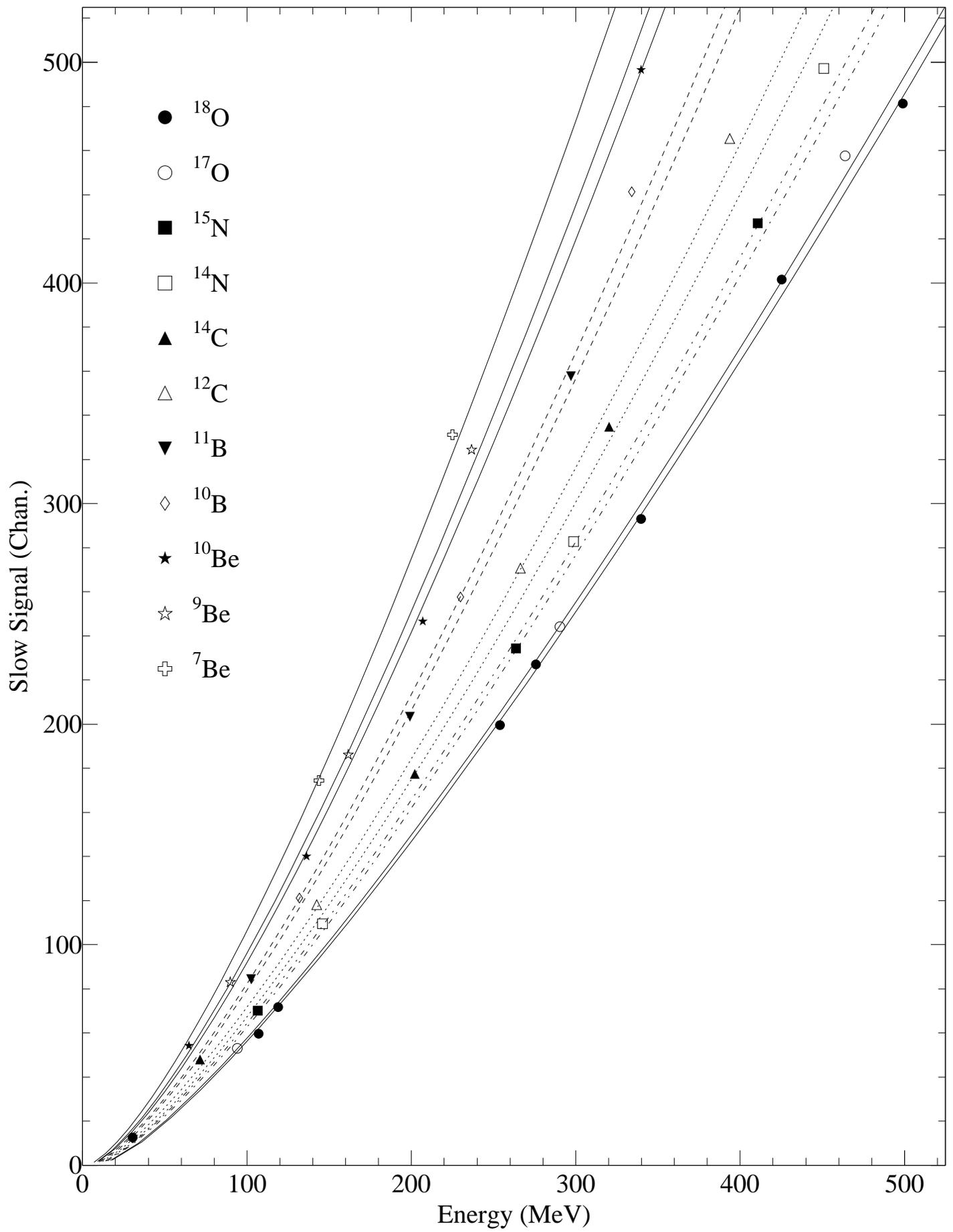
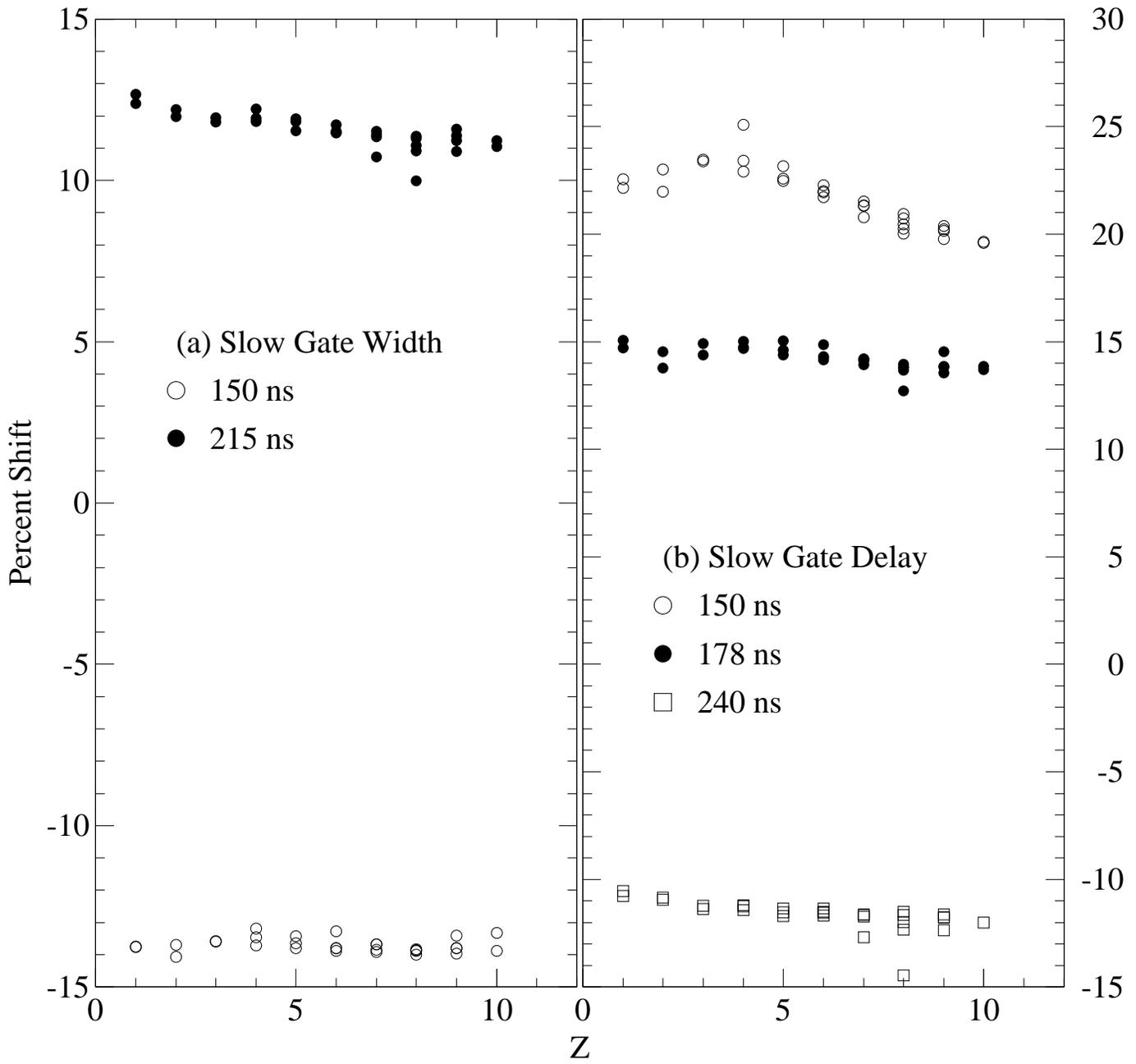
\end{document}